\documentclass[5p,twocolumn,12pt]{elsarticle}
\usepackage{amssymb}
\biboptions{sort&compress}
\usepackage{ifpdf}
\ifpdf
\usepackage[pdftex,
       colorlinks=true,
        pdftitle={Vortex patterns in a superconducting-ferromagnetic rod},
        pdfauthor={Antonio R. de C. Romaguera},
        pdfsubject={Vortex Matter in Nanostructured Superconductors (VORTEX VI)},
        pdfkeywords={Universidade Federal Rural de Pernambuco - UFRPE, Departamento de Física. Dom Manoel de Medeiros, s/n, Dois Irmãos, 52171-900, Recife-PE, Brasil, Tel: (55) 81-3320-6594},
        pdfcreator={antonio.romaguera@df.ufrpe.br},
        baseurl={http://www.df.ufrpe.br},
        pdfstartview=FitH,
        pdfpagemode=UseNone,
        bookmarksopen=true
        pdfoutput=1
        ]{hyperref}
\fi

\journal{Physica C}
\begin{document}
\begin{frontmatter}
\title{Vortex patterns in a superconducting-ferromagnetic rod}

\author[label1]{Antonio R. de C. Romaguera}
\author[label2]{Mauro M. Doria}
\author[label3]{Fran\c{c}ois M. Peeters}

\address[label1]{Departamento de F\'{\i}sica, Universidade Federal Rural de Pernambuco, \\52171-900 Recife, Pernambuco, Brazil}
\address[label2]{Departamento de F\'{\i}sica dos S\'{o}lidos, Universidade Federal do Rio de Janeiro, \\21941-972 Rio de Janeiro, Brazil}
\address[label3]{Departement Fysica, Universiteit Antwerpen, \\Groenenborgerlaan 171, B-2020 Antwerpen, Belgium}

\begin{abstract}
A superconducting rod with a magnetic moment on top develops
vortices obtained here through 3D calculations of the
Ginzburg-Landau theory. The inhomogeneity of the applied field
brings new properties to the vortex patterns that vary according to
the rod thickness. We find that for thin rods (disks) the vortex
patterns are similar to those obtained in presence of a homogeneous
magnetic field instead because they consist of giant vortex states.
For thick rods novel patterns are obtained as vortices are curve
lines in space that exit through the lateral surface.
\end{abstract}


\end{frontmatter}


\section{Introduction}
\label{Introduction}

The minimal condition for the onset of a vortex in a mesoscopic
superconductor depends on geometrical parameters.  A thin disk of
radius $R$ (thickness $D \sim \xi$, $\xi$ is the coherence length)
can only exist in the Meissner state for $R \sim \xi$. But for $\xi
< R < 2\xi$, giant vortex states are allowed and for $R
> 2\xi$ multivortex states become possible, as reported in
\cite{baelus01} and recently observed in Refs.
\cite{grigorieva:077005,kanda04}. Mesoscopic superconductors have
new and interesting properties\cite{romaguera07b}, and also provide
an interesting playground to understand the co-existence of
magnetism and superconductivity \cite{milo:052502,nature2004}. For
instance magnetic dots on top of superconducting film have been
investigated both theoretically  and experimentally
\cite{PhysRevB.59.14674,gheorghe:054502,aladyshkin:184519,Doria_romaguera_NP}.

In this paper we report a theoretical study done on superconducting
rods of radius $R$ and varying thickness $D$, with a magnetic dot on
top, a system which displays curved vortices triggered by the
inhomogeneity of the magnetic field. We address the minimal
geometrical conditions for the onset of vortices and also the nature
of the vortex patterns.

\begin{figure}[!b]
\centering
\includegraphics[width=0.4\linewidth]{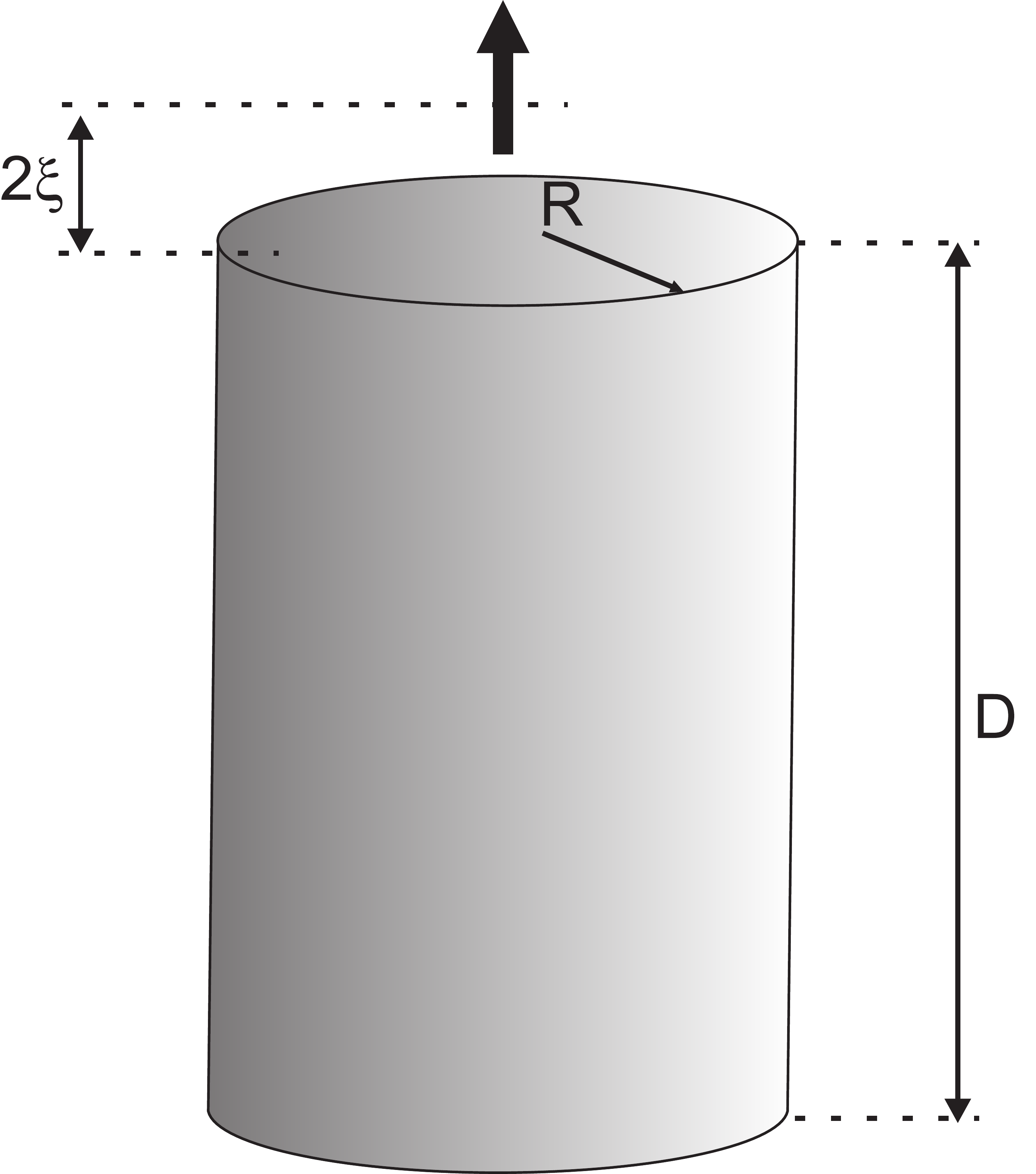}
\caption{(Color online) Superconducting rod with radius $R$ and
thickness $D$. The magnetic dot is oriented along the $x$-axis and
positioned $2\xi$ above the top surface.} \label{fig1}
\end{figure}

The shape of a vortex shifts from a flat "coin" to a curved line in
three-dimensional space, as the thickness $D$ tales the rod from a
disk to a tall rod. Consequently the vortex pattern consists of "top
to bottom" giant vortices and "top to side" multi-vortices in these
two extreme limits, namely, $D \sim \xi$ and $D \gg \xi$,
respectively. Rods of intermediate thickness display highly non
trivial vortex patterns showing features of both extreme limits. Our
results are obtained in the context of the Ginzburg-Landau theory in
the limit of no magnetic shielding. We take a magnetic dot with
magnetic moment $\mu$ positioned $2\xi$ above the rod's top surface
and oriented along the rod's axis, see Fig.~\ref{fig1}.

Although curve vortices also appear in presence of a homogeneous
field tilted with respect to the surface, as vortices must always
emerge perpendicular to the surface \cite{romaguera07}, the
inhomogeneity of the field brings new effects. The dipolar magnetic
field weakens with distance and may become so dim at the bottom of
the rod to the point that this region becomes unable to sustain
vortices. Consequently the Meissner state is kept at the bottom of
the rod. This favors a reentrant vortex state, as a top to side
vortex must disappear at the top surface, the same one of its onset.
Thus the Meissner phase is retrieved at a higher magnetic moment.
This qualifies superconducting rods with a magnetic moment on top
for technological applications as logical switchers, since vortices
can be expelled or accepted by fine tuning the external magnetic
moment dot strength.

\section{Theoretical approach}
\label{Theoretical approach}
According to the standard Ginzburg-Landau approach, the Gibbs free
energy of the superconductor near the critical temperature $T_{c}$
can be expanded in powers of the complex order parameter
$\Psi(\vec{r})$ that gives the density of Cooper-pairs:
$|\Psi(\vec{r})|^2$. Hence the Gibbs free energy difference between
the superconducting and the normal states is,
\begin{eqnarray*}
F_{s}-F_{n} = \int dV\; \bigg\{
\alpha(T) |\Psi|^2 + \frac{1}{2}\beta|\Psi|^4 + \nonumber \\
+\frac{\hbar^2}{2m^{*}}|\bigg(\vec{\nabla}-i\frac{2\pi}{\Phi_0}\vec{A}\bigg)\Psi|^2
\bigg\}\label{eq01}
\end{eqnarray*}
with the phenomenological constants $\alpha(T) \equiv
\alpha_0(T-T_c)<0 $, $\beta>0$, $m^{*}$ (the mass of one
Cooper-pair), and $\Phi_0=hc/2e$, the fundamental unit of flux.
Boundary conditions are imposed to the external surfaces of the rod.
We express quantities in this theory in dimensionless units defined
by the following reduced units: the coherence length
$\xi(T)=[-\hbar^2/2m^{*}\alpha(T)]^{1/2}$ (lengths),
$H_{c2}=\phi_0/2\pi\xi(T)^2$ (magnetic field),
$H_{c2}\xi=\Phi_0/2\pi\xi(T)$ (vector potential),
$\mu_0=H_{c2}\xi(T)^3=\Phi_0\xi(T)/2\pi$ (magnetic moment),
$\Psi_{0}=\sqrt{-\alpha(T)/\beta}$ (order parameter), and
$F_0=\alpha(T)^{2}/2\beta$ (free energy). The magnetization is
calculated using the expression $\vec M = \int \vec r \times \vec
J(r) dv$.

In terms of these dimensionless units, the Gibbs free energy difference becomes,
\begin{eqnarray}
&F&= 2\int dV\{-|\Psi|^{2} + \frac{1}{2}|\Psi|^{4} + \nonumber \\
&+& |(\vec{\nabla}-i\vec{A})\Psi|^{2}\},\label{free}\label{eq02}
\end{eqnarray}
The integration is restricted to the volume of the rod.
The boundary condition in dimensionless units becomes,
\begin{equation} \label{eq03}
\vec{n}\cdot (\vec{\nabla}-i\vec{A})\Psi \bigg|_{boundary}=0.
\end{equation}
%
For the magnetic dot we use field produced by a point-like dipole $\vec A= (\vec \mu \times \vec r)/r^3$.

\begin{figure}[!ht]
\centering
\includegraphics[width=0.8\linewidth]{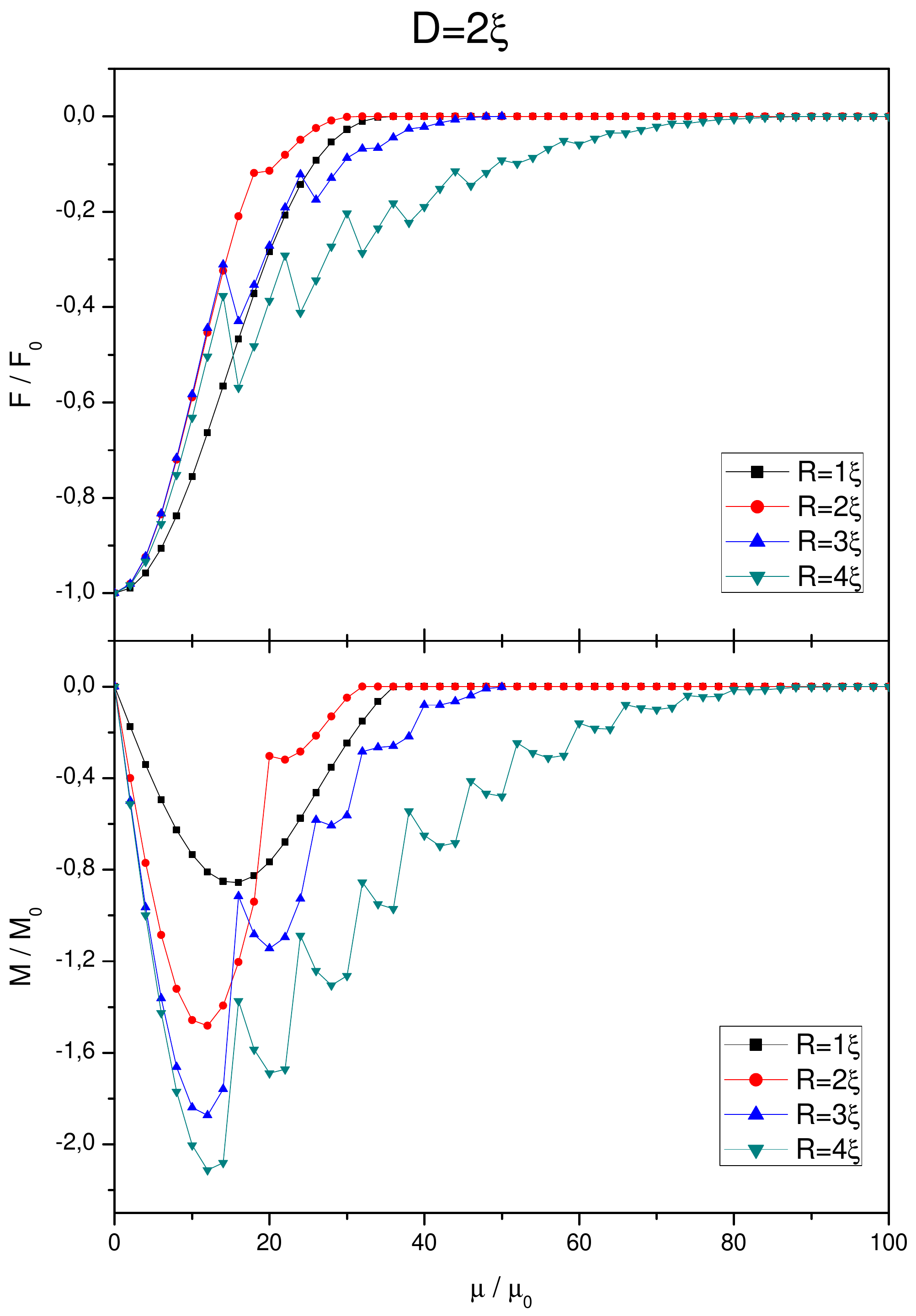}
\caption{(Color online) Free energy $F/F_0$ and magnetization
$M/M_0$ versus magnetic moment $\mu/\mu_0$ for mesoscopic rods with
thickness $D=2\xi$ and several radii. Giant vortices with increasing
maximum vorticity are observed by increasing the radius.}
\label{fig2}
\end{figure}

\section{Numerical results}
\label{Numerical results}
We solved Eqs.~\ref{eq02} and \ref{eq03} using the simulated
annealing procedure \cite{romaguera07b}. We set the following
parameters for the rod: the radius $R$ ranges from $1\xi$ to $4\xi$;
the thickness $D$ ranges from $2\xi$ to $8\xi$ and the magnetic
moment $\mu$ ranges from $0$ to $100\mu_0$.  For the thinnest rod
considered here, $D=2\xi$, we find that the inhomogeneity of the
field does not matter, since the free energy and the magnetization
are similar to the homogeneous applied field case. Clearly this is
because the dot's magnetic field does not vary significantly inside
the rod.

For $R=1\xi$ and any thickness we find no vortex state, thus the
Meissner state prevails up to the normal state. We do not observe
anti-vortices or Meissner state with opposite orientation in the
rods because we limit the present study to rods with a maximum
radius of $4\xi$. Thus only giant vortex states (GVS) are observed.
For $R=2\xi$ and for $D=2\xi$, only one vortex is allowed. For
larger $R$ more than one vortex is possible and their number
increases very rapidly with the radius. Although there is more space
to accommodate extra vortices the magnetic field becomes weaker at
distances away from the center. Consequently the average magnetic
magnetic field does not grow with the radius. This effect is
equivalent to that studied by Ovchinnikov sometime ago
\cite{Ovchinnikov}. A sawtooth behavior in the magnetization with
bigger $H_{c2}$ is clearly established for bigger radius. This
behavior is observed for $R=3\xi$ and $D=2\xi$. In Fig.~\ref{fig2}
we show the vortex states obtained for different thin rods, i.e.,
disks. Their magnetization curves are similar to those obtained in
presence of homogeneous field. In table \ref{table1} we give the
complete description of the states through the set of parameters
($R,D,\mu/\mu_0$).

\begin{figure}[!b]
\centering
\includegraphics[width=0.85\linewidth]{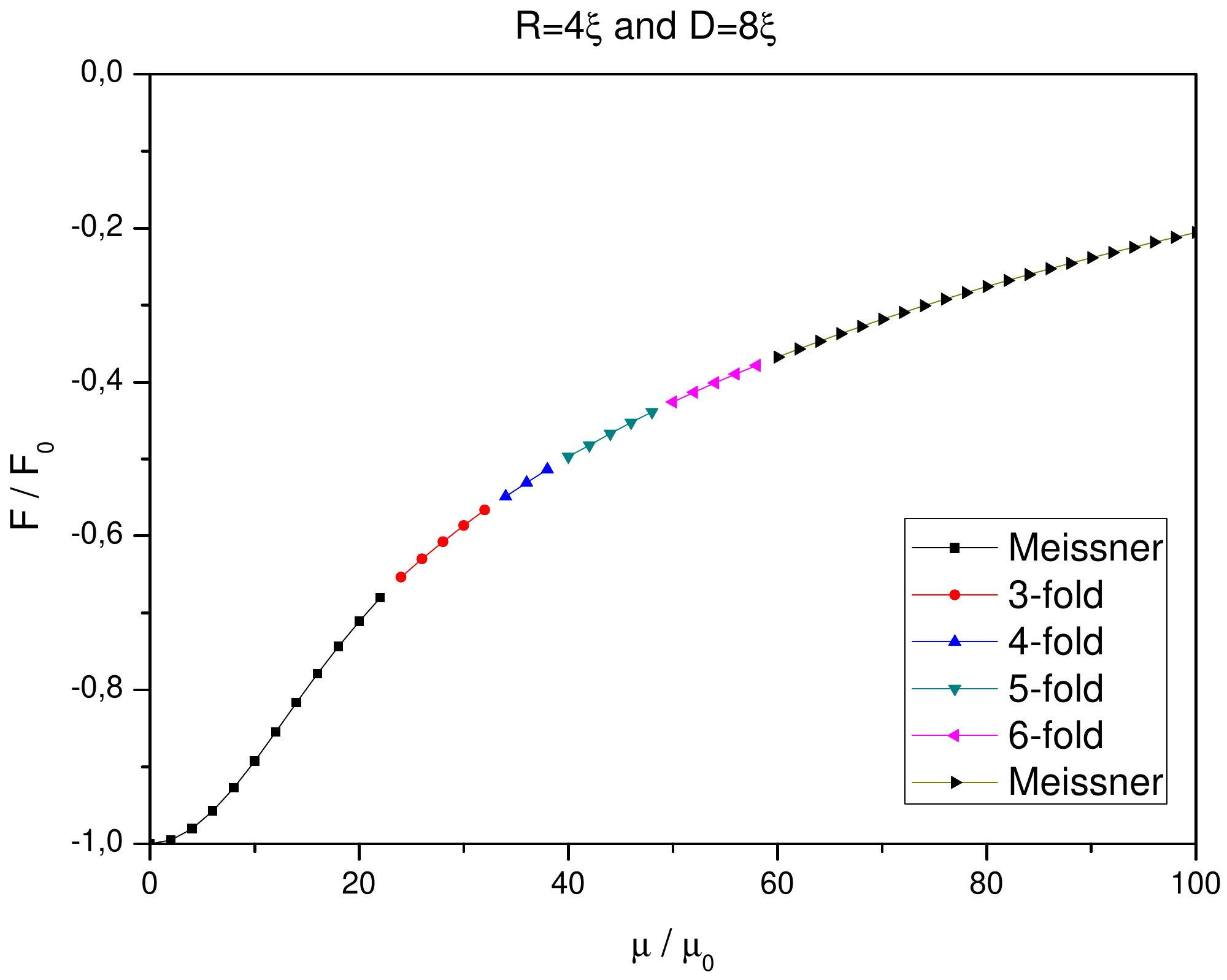}
\caption{(Color online) Free energy $F/F_0$ vs $\mu/\mu_0$ for a rod
with $R=4\xi$ and $D=8\xi$. For increasing magnetic moment the
vortex pattern evolves continuously through the N-fold
configurations, as labeled in the inset. See, Fig.~\ref{fig4} for
the corresponding $|\psi|^2$ isosurface views of these N-fold
states.} \label{fig3}
\end{figure}

The inhomogeneity of the field becomes important for the thicker
rods $D=6\xi$ and $8\xi$, as shown here. Even for small radii, the
rod already displays behavior that differs significantly from the
homogeneous field case. We find smooth free energy lines, with no
first order transition occasioned by the entrance of quantized
vortex lines. Fig.~\ref{fig3} shows the free energy for the $R=4\xi$ and $D=6\xi$
rod.

\begin{figure}[ht]
\centering
\includegraphics[width=0.85\linewidth]{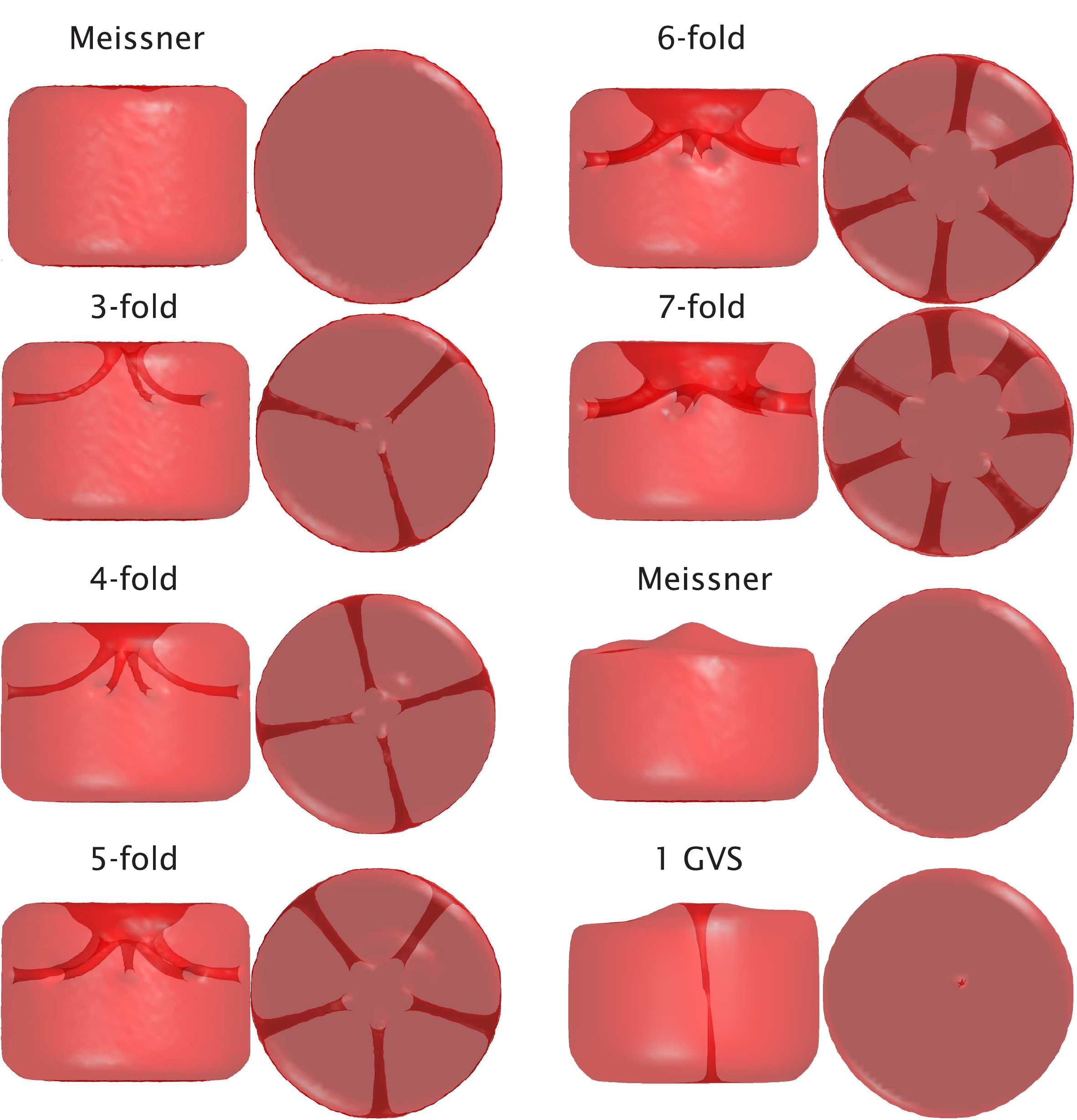}
\caption{(Color online) Isosurfaces of the Cooper pair density for a mesoscopic rod with $R=4\xi$ and $D=6\xi$. Every image corresponds to one of the states N-folder states listed in Table \ref{table1}. }
\label{fig4}
\end{figure}

\begin{table*}[t]
\caption{The sequence of ground states for the of rods considered here. The magnetic dot is positioned $2\xi$ away from the rod's top surface and oriented along the $z$-axis. The question mark means the normal state was not reach.}
\centering
\tiny
\begin{tabular}{ccccccccccccccccc}
\hline
\hline
$R/\xi$&1&1&1&1&2&2&2&2&3&3&3&3&4&4&4&4\\
$D/\xi$&2&4&6&8&2&4&6&8&2&4&6&8&2&4&6&8\\
State&$\mu/\mu_0$&$\mu/\mu_0$&$\mu/\mu_0$&$\mu/\mu_0$&$\mu/\mu_0$&$\mu/\mu_0$&$\mu/\mu_0$&$\mu/\mu_0$&$\mu/\mu_0$&$\mu/\mu_0$&$\mu/\mu_0$&$\mu/\mu_0$&$\mu/\mu_0$&$\mu/\mu_0$&$\mu/\mu_0$&$\mu/\mu_0$\\
\hline
Meissner&0-36&$<100$&$<100$&$<100$&0-18&0-76&$<$100&$<$100&0-14&0-46&$<$100&$<$100&0-14&0-20&0-22&0-22\\

1 GVS&-&-&-&-&20-30&78-100&-&-&16-24&48-76&-&-&16-22&22-100&-&-\\

2 GVS&-&-&-&-&32&-&-&-&26-30&78-100&-&-&24-30&-&-&-\\

3 GVS&-&-&-&-&-&-&-&-&32-38&-&-&-&32-36&-&-&-\\

4 GVS&-&-&-&-&-&-&-&-&40-50&-&-&-&38-44&-&-&-\\

5 GVS&-&-&-&-&-&-&-&-&-&-&-&-&46-50&-&-&-\\

3-fold&-&-&-&-&-&-&-&-&-&-&-&-&-&34-38&24-34&24-32\\

4-fold&-&-&-&-&-&-&-&-&-&-&-&-&-&40-44&36-42&34-40\\

5-fold&-&-&-&-&-&-&-&-&-&-&-&-&-&46-52&44-48&42-48\\
6-fold&-&-&-&-&-&-&-&-&-&-&-&-&-&54-56&50-54&50-58\\
7-fold&-&-&-&-&-&-&-&-&-&-&-&-&-&-&56-62&-\\
Meissner&-&-&-&-&-&-&-&-&-&-&-&-&-&-&64-74&60-100\\
1 GVS&-&-&-&-&-&-&-&-&-&-&-&-&-&-&76-100&-\\
Normal&$>36$&?&?&?&$>32$&?&?&?&$>50$&?&?&?&?&?&?&?\\
  \hline
  \hline
\end{tabular}
\label{table1}
\end{table*}
%
%
The 3D view of the Cooper pair density, $|\psi|^2$, shows curved
top-to-side vortices, that is, vortices that enter from the top
surface and exit perpendicularly in the side surface. We call them
N-fold states, where N describes the number of vortices. Along the
energy curve vs $\mu$ in Fig.~\ref{fig3} the states are labeled
according to this definition. The N-fold states are not clearly seen
in Fig.~\ref{fig3} because their free energy lines is superposed to
the Meissner line from where is their onset and disappearance.
However these states are clearly seen to exist by analyzing the
$|\psi|^2$ isosurface, as shown in Fig.~\ref{fig4}. In the two
biggest rods considered here we observed the nucleation of N-fold
states followed by Meissner and GVS. This features are justified by
the growth of a normal region closer to the the magnetic dot
position. When this normal region occupies a significant part of the
rod the remaining superconducting part behaves as a thin rod, i.e.,
it does not exhibits N-fold states. So, we retrieve the Meissner
state. This feature is in Table~\ref{table1} and correspond to the
second Meissner state.

\section{Conclusions}
\label{Conclusions}

Using Simulated Annealing, a truly three-dimensional approach, we
obtain the vortex patterns of mesoscopic rods with a point-like
magnetic moment on top. We conclude that rods with thickness smaller
than $4\xi$ can be considered as thin disks since only top to bottom
giant vortex states are obtained.  For rods with thickness bigger
than $4\xi$, we observed the appearance of N-fold vortex states,
namely N top-to-side vortices. The onset and disappearance of these
states is from a single GVS or Meissner state line, as continuous
transitions from it. Consequently, for sufficient large magnetic
moments one can retrieve the Meissner state before reaching the
normal state, and this, leads to a reentrant behavior.

\section{Acknowledgment}
\label{Acknowledgment}


A. R. de C. Romaguera acknowledges the brazilian agency FACEPE for
financial support. M. M. Doria acknowledges CNPq and FAPERJ. F. M.
Peeters acknowledges Flemish Science Foundation (FWO-Vl), the
Belgian Science Policy (IUAP) and the ESF-AQDJJ network.

\bibliographystyle{elsarticle-num}
\bibliography{reference_vortex}

\begin{thebibliography}{10}
\expandafter\ifx\csname url\endcsname\relax
  \def\url#1{\texttt{#1}}\fi
\expandafter\ifx\csname urlprefix\endcsname\relax\def\urlprefix{URL }\fi
\expandafter\ifx\csname href\endcsname\relax
  \def\href#1#2{#2} \def\path#1{#1}\fi

\bibitem{baelus01}
B.~J. Baelus, F.~M. Peeters, V.~A. Schweigert, Phys. Rev. B 63~(14) (2001)
  144517.

\bibitem{grigorieva:077005}
I.~V. Grigorieva, W.~Escoffier, J.~Richardson, L.~Y. Vinnikov, S.~Dubonos,
  V.~Oboznov, Phys. Rev. Lett. 96~(7) (2006) 077005.

\bibitem{kanda04}
A.~Kanda, B.~J. Baelus, F.~M. Peeters, K.~Kadowaki, Y.~Ootuka, Phys. Rev. Lett.
  93~(25) (2004) 257002.

\bibitem{romaguera07b}
A.~R. de~C.~Romaguera, M.~M. Doria, F.~M. Peeters, Phys. Rev. B 76~(2) (2007)
  020505.

\bibitem{milo:052502}
M.~V. Milo\v{s}evi\'{c}, G.~R. Berdiyorov, F.~M. Peeters, Phys. Rev. B 75~(5)
  (2007) 052502.

\bibitem{nature2004}
Z.~Yang, M.~Lange, A.~Volodin, R.~Szymczak, V.~V. Moshchalkov1, Nature Material
  3 (2004) 793.

\bibitem{PhysRevB.59.14674}
M.~J. Van~Bael, K.~Temst, V.~V. Moshchalkov, Y.~Bruynseraede, Phys. Rev. B
  59~(22) (1999) 14674.

\bibitem{gheorghe:054502}
D.~G. Gheorghe, R.~J. Wijngaarden, W.~Gillijns, A.~V. Silhanek, V.~V.
  Moshchalkov, Phys. Rev. B 77~(5) (2008) 054502.

\bibitem{aladyshkin:184519}
A.~Y. Aladyshkin, D.~A. Ryzhov, A.~V. Samokhvalov, D.~A. Savinov, A.~S.
  Mel'nikov, V.~V. Moshchalkov, Phys. Rev. B 75~(18) (2007) 184519.

\bibitem{Doria_romaguera_NP}
M.~M. Doria, A.~R. de~C.~Romaguera, F.~M. Peeters, Unpublished.

\bibitem{romaguera07}
A.~R. de~C.~Romaguera, M.~M. Doria, F.~M. Peeters, Phys. Rev. B 75~(18) (2007)
  184525.

\bibitem{Ovchinnikov}
Y.~N. Ovchinnikov, Sov. Phys. JETP 52 (1980) 775.

\end{thebibliography}

\end{document}